# Compensated magnetic insulators for extremely fast spin-orbitronics

Heng-An Zhou[1,2], Yiqing Dong[1,2], Teng Xu[1,2], Kun Xu[3], Luis Sánchez-Tejerina[4], Le Zhao[1,2], You Ba[1,2], Pierluigi Gargiani[5], Manuel Valvidares[5], Yonggang Zhao[1,2], Mario Carpentieri[4], Oleg A. Tretiakov[6], Xiaoyan Zhong[3], Giovanni Finocchio[7], Se Kwon Kim[8], and Wanjun Jiang[1,2,†]

[1]*State Key Laboratory of Low-Dimensional Quantum Physics and Department of Physics, Tsinghua University, Beijing 100084, China*
[2]*Frontier Science Center for Quantum Information, Tsinghua University, Beijing 100084, China*
[3]*Key Laboratory of Advanced Materials (MOE), School of Materials Science and Engineering, Tsinghua University, Beijing 100084, China*
[4]*Department of Electrical Engineering and Information, Politecnico di Bari, Bari 70125, Italy*
[5]*ALBA Synchrotron Light Source, Cerdanyola del Vallès, 08290 Barcelona, Spain*
[6]*School of Physics, The University of New South Wales, Sydney 2052, Australia*
[7] *Department of Mathematical and Computer Sciences, Physical Sciences and Earth Sciences, University of Messina, Messina 98166 Italy*
[8]*Department of Physics and Astronomy, University of Missouri, Columbia, MO 65211, USA*

[†] To whom correspondence should be addressed:
jiang_lab@tsinghua.edu.cn

**The fast spin dynamics provide many opportunities for the future communication and memory technologies[1-5]. One of the most promising examples is the domain wall (DW) racetrack memory[1,6-9]. To achieve fast device performances, the high-speed motion of DWs[10-13] is naturally demanded that leaves antiferromagnets (AFMs) and compensated ferrimagnets (FIMs) as the promising materials. While controlling and probing the dynamics of DWs in AFMs remains challenging, the fast motion of DWs with velocities around 1500 m/s has been demonstrated in metallic FIMs. The velocity of DWs in metallic FIMs is, however, suppressed by the magnetic damping of conduction electrons, which motivates us to explore how fast DWs can move in insulating FIMs where the conduction electron is absent. In this work, through synthesizing compensated FIM insulator $Gd_3Fe_5O_{12}$ thin films with a perpendicular magnetic anisotropy, we demonstrate that the spin-orbit torque (SOT) induced motion of DWs along the $Gd_3Fe_5O_{12}$/Pt racetrack can approach 6000 m/s. Our results show that the exceptionally fast motion of DWs can be achieved in compensated FIM insulators owing to small damping inherent to magnetic insulators, which could potentially facilitate the emerging ultrahigh-speed spintronic logics and racetrack memories by introducing insulating compensated FIMs as a new material platform.**

The achievement of high-speed motion of DWs in AFMs is key to realizing the future ultrafast spintronic memory and logic devices[1-3,6-20], which, however, has been elusive so far. This can be attributed to the experimental difficulties in controlling and probing the dynamics of DWs in AFMs due to the lack of net magnetization and their resultant weak electrical/optical responses[10,12,15-20]. By contrast, compensated metallic FIMs has emerged as an alternative platform for demonstrating the fast motion of DWs[11-13]. In this work, we will explore the possibility of achieving even faster motion of DWs in fully compensated FIM insulators, $Gd_3Fe_5O_{12}$ (GdIG), which is made possible by prohibiting the extra spin damping from conduction electrons. In GdIG thin films, the magnetic moments of the $Gd^{3+}$ ions are antiferromagnetically coupled with the magnetic moments of $Fe^{3+}$ ions[1,21]. GdIG has two special temperatures due to the fact that the two magnetic elements have different Landé factors[11-13,22,23], $g_{Fe} \sim 2.2$ and $g_{Gd} \sim 2$: the magnetization compensation temperature $T_M$, where the net magnetization vanishes $M_{Fe} = M_{Gd}$ and the angular momentum compensation temperature $T_A$ where the net spin density vanishes $S_{Fe} = S_{Gd}$. The zero-magnetization at $T_M$ and the zero-spin density at $T_A$ can give rise to the AFM-like dynamics in GdIG, in comparison with other frequently studied FIM garnets such as $Y_3Fe_5O_{12}$ and $Tm_3Fe_5O_{12}$[1,24-27], where complete magnetization compensation is absent. Meanwhile, due to the relatively low Fermi level of Gd and weak coupling between electron spins and magnetic moments in the inner $4f$ shells of Gd, electrical and optical measurements are more sensitive to the magnetization of Fe in GdIG. Thus, using standard laboratory based electrical transport and magneto-optic Kerr effect (MOKE) imaging techniques, the dynamics of fully compensated DWs in FIM GdIG can be readily investigated, which in turn, provide rich insights for designing the future ultrafast DW-based spintronic devices[11-13].

The statics and dynamics of a compensated Néel-type FIM DW in GdIG/Pt bilayers can be obtained numerically and analytically from solving two coupled Landau-Lifshitz-Gilbert (LLG) equations including the SOT contributions[7,11,28,29], as discussed in the Supplemental Materials. Considering an effective model (see Supplemental Materials for the link between the two LLG equations and the model parameters), the resultant steady-state velocity is given by:

$$V = \frac{\pi q}{2} \cdot \frac{D j_e}{\sqrt{(\delta_s j_e)^2 + \left(\frac{2e\alpha s t_f D}{\hbar \Delta \theta_{sh}}\right)^2}} \qquad (1)$$

where $q = \pm 1$ represents the type of the Néel DW (the magnetization direction changes from $q\hat{z}$ to $-q\hat{z}$ as $x$ increases), $D$ is the Dzyaloshinskii-Moriya interaction (DMI) coefficient, $j_e$ is the charge current flowing along the $\hat{x}$ direction, $\delta_s = S_{Fe} - S_{Gd}$ is the net spin density, $s = S_{Fe} + S_{Gd}$ is the saturated spin density, $\alpha = (\alpha_{Fe} S_{Fe} + \alpha_{Gd} S_{Gd})/s > 0$ is the damping constant, $e$ is the electron charge, $\hbar$ is the reduced Planck constant, $\Delta$ is the DW width, $t_f$ is the thickness of the FIM layer, and $\theta_{sh}$ is the spin Hall angle of Pt. When temperature is far away from $T_A$ and the current density $j_e$ is sufficiently large (namely, $\delta_s j_e \gg 2e\alpha s t D/\hbar \Delta \theta_{sh}$), the DW velocity saturates as $V \cong \pi q|D| \operatorname{sgn}(j_e)/2|\delta_s|$. This is identical to the case of FM where the saturation of DW motion is given by $V \cong \gamma \pi q|D|\operatorname{sgn}(j_e)/2M_s$ with $\gamma$ being the gyromagnetic ratio and $M_s$ the saturation magnetization. Thus, the DW velocity in FIMs is generally faster as a result of their reduced net spin density $\delta_s$. In particular, $Eq.(1)$ indicates that the DW velocity $V$ attains its maximum when $\delta_s = 0$, $i.e.$, at $T_A$:

$$V = \frac{\pi q}{2} \cdot \frac{\hbar}{2e} \cdot \frac{\Delta \theta_{sh}}{\alpha s t_f} \cdot j_e \qquad (2)$$

Clearly, the DW velocity at $T_A$ increases without saturation as the current density $j_e$ increases (provided that the current-induced change of DW width $\Delta$, which ultimately limits the DW velocity below the maximum spin-wave group velocity[28,30], is small enough to be neglected), which is in stark contrast with the frequently studied FM systems.

Following the above theoretical discussions, various compensated FIM GdIG films with thicknesses ranging from $t_f$ = 7 nm to $t_f$ = 30 nm have been epitaxially grown onto (111)-oriented $Gd_3Sc_2Ga_3O_{12}$ (GSGG) substrates. Epitaxial growth and atomically sharp interface are confirmed by cross-sectional scanning transmission electron microscopy (STEM), as shown in Fig. 1A. The lattice mismatch (-0.67%) between the substrate ($a$ = 12.554 Å) and GdIG films ($a$ = 12.471 Å) induces a perpendicular magnetic anisotropy (PMA) in the GdIG films[31,32]. Magnetic hysteresis loops at different temperatures (165-300K) are also measured, revealing a vanishing net saturation magnetization $M_s$ around 220 K, and $M_s$ = 33 emu/cc (300 K), as shown in Fig. 1B. More structural and magnetization characterization data can be found in the Figs.S1-S3 of the Supplementary Materials.

The electrical manipulation of magnetic insulators is made possible by harvesting the strong spin-orbit interaction of heavy metal Pt layer in proximity with insulating GdIG films. Specifically, in GdIG/Pt bilayers, the magnetization dynamics of the (insulating) GdIG films can be electrically probed (via spin Hall magnetoresistance - SMR)[31,33] and manipulated (via current-induced SOTs)[32,34-36], through the (inverse) spin Hall effects of the (conducting) Pt layers. Fig. 1C shows the SMR signals measured in the Hall configuration ($R_{xy}^{SMR-AHE}$) as a function of the perpendicular magnetic field ($H_z$). The sign reversal of $R_{xy}^{SMR-AHE}$ loops is associated with the magnetization compensation temperature ($T_M \sim$ 220 K), across which the magnetizations of the AFM-coupled $M_{Fe}$ and $M_{Gd}$ reverse sign accordingly. The coercive fields ($H_c$) and $M_s$ are summarized in Figs. 1(D)-1(E), enhancement of $H_c$ and vanishing of $M_s$ are observed simultaneously in the vicinity of $T_M$ which is consistent with the enhanced AFM-coupling between $M_{Fe}$ and $M_{Gd}$ ions. The presence of AFM-coupling between Fe and Gd elements were further confirmed through X-ray magnetic circular dichroism (XMCD) measurements. Shown in Figs. 1F-1I are element-specific magnetic hysteresis loops for Fe and Gd across $T_M$, respectively. At 175 K and 250 K, the opposite hysteresis loops confirm the onset of AFM-coupling between 3$d$ Fe and 4$f$ Gd elements. Different $T_M$ between the bulk samples ($T_M \sim$ 291 K) and the present thin films ($T_M \sim$ 220 K) can be explained by the selective sputtering of compound targets and the oxygen stoichiometry in the GdIG films. Further, deterministic switching of perpendicular magnetization vectors by current-induced

SOTs in the (15nm)GdIG/(4nm)Pt bilayers is demonstrated and shown in the Fig. S6 of the Supplementary Materials, which confirms the feasibility of manipulating compensated DWs in the present insulating FIM system.

The subsequent motion of compensated DWs in the GSGG/(15 nm)GdIG/(4 nm)Pt trilayer is studied using racetrack-like devices shown in Fig. 2A. In our device, electrical currents in the conducting Pt layer moves the compensated DWs in insulating GdIG films through (damping-like) SOTs generated by the spin Hall effects in Pt[36]. Figs. 2B-2C show series of polar MOKE images acquired in the presence of $\pm H_z$, after applying consecutive pulse (electron) currents with amplitude $j_e = 1.08 \times 10^8$ A/cm$^2$ and duration 28 ns. The (average) velocity of DWs ($V = 50$ m/s) is calculated after dividing the measured displacements by pulse durations. The direction of DW motion is reversed after inverting $H_z = +8$ mT to $H_z = -8$ mT, as shown in Figs. 2B-2C. The response of Néel or Bloch DWs driven by SOTs can be understood from the symmetry of the effective spin Hall field[8,9,37], $\vec{H}_{sh} = H_{sh}^0 (\hat{m} \times (\hat{z} \times \hat{j}_e))$. Here $\hat{m}$ denotes the orientation of the central spin inside the DW (handedness of DWs), $\hat{z}$ is the unit vector along the $z$ direction and $\hat{j}_e$ is the electron current direction. The prefactor $H_{sh}^0$ governs both the efficiency and sign of SOTs ($H_{sh}^0 > 0$, since $\theta_{sh} > 0$ for Pt). The symmetry of $\vec{H}_{sh}$ indicates that only Néel DWs move by SOTs, whereas Bloch DWs remain stationary.

Similar with other interfacially symmetric multilayers[4,5,38], the spin chirality is governed by the DMI at the GSGG/GdIG and the GdIG/Pt interfaces[24,26]. Based on the SOT magnetometry[24,39,40], strengths of the interfacial DMI in GSGG/GdIG/Pt trilayers were estimated as $D \approx 1.9$ μJ/m$^2$ for 15 nm GdIG film and $D \approx 3.4$ μJ/m$^2$ for 7 nm GdIG film, respectively, as discussed in the Fig. S7 of the Supplementary Materials. Note that while thinner GdIG films (7-15 nm) exhibit both PMA and larger interfacial DMI, their polar MOKE contrasts are however, beyond the sensitivity of our microscope, and the motion of compensated DWs thus cannot be detected in the thinner GdIG films. These DMI values are consistent with reported results from GSGG/TmIG(10 nm)/Pt trilayers[24,26], in which the existence of mixed Bloch and Néel type spin structures inside DWs is suggested[26]. This could explain the contraction of domains in the presence of a negative $H_z$, as a result of the same direction of $\vec{H}_{sh}$ exerted on the achiral Néel-type DWs (↑↗→↘↓↘→↗↑).

Following the increase of current densities, we have observed the motion of DWs with a velocity $V = 2950 \pm 30$ m/s at a current density $j_e = 1.08 \times 10^8$ A/cm$^2$ (duration 34 ns) and at $H_z = +8$ mT, as shown in Fig. 2D. DW velocity as a function of pulse amplitudes and durations are summarized in Fig. 2E. For $j_e < 1 \times 10^8$ A/cm$^2$, the velocity of DWs ($<$ 50 $m/s$) increases linearly with $j_e$ and does not depend on the pulse duration. Upon approaching $1 \times 10^8$ A/cm$^2$, the velocity is significantly increased. Shown in Fig. 2F is the evolution of velocity as a function of pulse duration at $j_e = 1.08 \times 10^8$ A/cm$^2$, which shows a similar nonlinear increase of velocity following the increase of durations. Fig. 2G corresponds to the nonlinear dependence of DW velocity as a function of $H_z$. This nonlinear behavior of velocity, may be due to the thermal weakening of PMA that produces a transient multidomain state during motion, together with the inertia effects of DWs moving at an extremely high speed. However, uncovering the origin of this nonlinearity needs further investigation. When durations are longer than 35 ns, DWs move out of the field of view of microscope after applying one single pulse. Thus, the corresponding velocities $V$ which are presumably higher than 3200 m/s, cannot be precisely measured. It is also conceivable that the DW velocity can be faster upon applying shorter pulse currents with bigger amplitudes that is however, limited by the range of the present instrument. All raw MOKE imaging data were presented in the Part 6 of the Supplementary Materials. Note that the maximum DW velocity $V_{max}$ can be estimated based on the maximum spin-wave group velocity $V_{max} \sim 2A/(d \cdot s) \sim 10000$ m/s, where $A$ = 3.2 pJ/m is the exchange coefficient, $d \approx 1.2$ nm is the lattice constant and $s$ is the spin density.

The spin structures inside DWs can be tuned to be either Néel or Bloch types when the applied in-plane fields ($H_{x,y}$) are larger than the DMI effective field $H_D$ [8,9,39]. Therefore, we subsequently studied the dynamics of Néel and Bloch DWs at room temperature, as shown in Fig. 3. For sufficiently strong magnetic fields applied in the $\hat{x}$ direction ($H_x$), the SOT-induced DW velocity is given by *Eq.* (1) with the replacement of $D$ by $M_s H_x$ (See the Supplemental Information for the further details). Therefore, the DW velocity is expected to change its sign when the sign of the following quantities is reversed: the current $\pm j_e$, the external fields $\pm H_x$, and the type of Néel DW $q = \pm 1$, which was confirmed by our experiments as shown in Fig. 3A. For $H_x = \pm 40$ mT, the direction of DW motion is reversed upon inverting the polarity of pulse currents accordingly, with a maximum velocity $V \approx$ 3600 $\pm 100$ m/s at $j_e = \pm 1.08 \times 10^8$ A/cm$^2$ (duration 25 ns), as shown in Fig. 3A. Shown in the sets are MOKE snapshots acquired before and after applying a single pulse currents (25 ns), through which the velocity is calculated. Directions of DW motion are also consistent with the symmetry of $\vec{H}_{sh}$ for different types of Néel DWs, as schematically illustrated in the inset. More data acquired at different pulse durations and current densities are summarized in Figs. 3B and 3C. The increase of $H_x$ give a more stable Néel type DW leading to a significantly enhanced of DW velocity, as implied by Fig. 3A. In particular, we have observed the remarkably fast motion of DW with a velocity $|V| = 5800 \pm 200$ m/s at $H_x = \pm 63$ mT, driven by a single pulse of duration 18 ns and of amplitude $j_e = \pm 1.08 \times 10^8$ A/cm$^2$, which can be found in Fig. S16 of the Supplementary Materials. Importantly, unlike the saturation of DW velocity in FMs, the experimentally observed velocity of compensated DWs in FIMs does not show trend towards saturation. In contrast, after applying magnetic fields along $y$ axis ($H_y$) that alters DWs from Néel type to Bloch type, no motion of DWs is observed after applying $j_e = \pm 1.08 \times 10^8$ A/cm$^2$ (18 ns), as shown in Fig. 3D, which is expected since $\vec{H}_{sh} = 0$ for Bloch-type DWs[7-9,37,39]. Note that the same current density could induce the motion of Néel-type DWs with a velocity $V = 5800$ m/s, which allows us to exclude current-induced thermal effect as the possible origin responsible for the observed fast motion of DWs. Based on the two coupled LLG equations, micromagnetic simulation studies for compensated DWs driven by SOTs were also carried out[41], which confirm qualitatively our experimental results discussed above, as given in the Part 8 of the Supplementary Materials.

The velocity of DWs should be substantially enhanced in the vicinity of angular momentum compensation temperature $T_A$, as theoretically predicted above. To fully track the motion of compensated DWs in the field of view of microscope, we adopted a relatively smaller current density ($j_e = 1 \times 10^8$ A/cm$^2$) at different temperatures. Following the decrease of temperatures, an enhanced DW velocity up to $V = 3500 \pm 300$ m/s is evident at $T = 230$ K (pulse duration 35 ns). This observation thus validates the theoretical prediction given by $Eq.$ (2). Note that external magnetic fields $H_z$ were normalized by the corresponding coercive fields $H_c$ for each measured temperature. The magnetization compensation temperature $T_M$ is shown in Fig. 4B, in which a vanishing $M_S$ is identified at $T_M = 220$ K. Note that the maximum DW velocity occurs at $T = 230$ K, instead of $T_M = 220$ K, which thus enables the angular momentum compensation temperature to be identified as $T_A = 230$ K. Furthermore, the evolution of DW velocity driven by varying duration of pulse currents across $T_A$ is also studied. Similar features as those shown in Fig. 4A are provided in the Fig. S18 and reproduced theoretically by micromagnetic simulations in the Part 13 the Supplementary Materials. Note that the velocity of DWs around $T_A$ could be even faster upon applying shorter pulses with larger amplitudes, in which influences of thermal weakening of PMA by current-induced Joule heating is less pronounced[13]. For potential applications, a prototypical DW racetrack-like device is constructed for implementing the fast motion of compensated DWs, as shown in Fig. 4B. In our GdIG/Pt bilayer devices, the presence and absence of DW can be electrically detected via using the SMR measured in the anomalous Hall configuration, in which the high/low SMR voltages $V_{xy}$ represent 1 and 0, respectively.

The exceptionally fast motion of compensated DWs approaching 6000 m/s has been demonstrated at room temperature in racetrack devices made of $Gd_3Fe_5O_{12}$/Pt bilayers, which occurs as a consequence of small effective damping coefficient $\alpha_{eff} = 0.04$ due to the absence of spin-flip scattering of conduction electrons, as compared with other metallic FIMs[22]. By integrating the SOT-enabled extremely fast motion of compensated DWs with electrical detection methods, one could benefit from the ultrahigh-speed DW racetrack devices. Similarly, by synthesizing interfacially asymmetric epitaxial heterostructures, one could also harvest the enhanced interfacial DMI in insulating multilayers[5]. This could lead to the stabilization of AFM-like topological spin textures, which should exhibit a much lower threshold driving current and many novel topological phenomena[20,38,42]. Thus, a comprehensive understanding of the extremely fast motion of the compensated DWs in insulating FIMs could largely promote the future ultrahigh-speed and energy efficient spintronics.

**Methods**

Thin-film FIM of nominal composition $Gd_3Fe_5O_{12}$ were grown by using an AJA ultrahigh vacuum magnetron sputtering system onto (111)-oriented scandium substituted gadolinium gallium garnet ($Gd_3Sc_2Ga_3O_{12}$, GSGG) substrates at 750 ℃ in an oxygen-rich environment and naturally cooled down to room temperature. GSGG substrates were firstly treated in $O_2$ environment at 1000 ℃ for 6 hours before loading into the chamber for deposition. This step is necessary for achieving atomically sharp surfaces and for ensuring subsequent high-quality epitaxial growth of $Gd_3Fe_5O_{12}$ films. For studying SMR and current-induced SOTs, a 4-nm thick Pt layer was deposited at a rate of 0.2 Å/s at room temperature without breaking the ultrahigh vacuum of the main chamber, which is necessary to demonstrate the spin-orbit detection/manipulation. The (111)-oriented GSGG substrates and $Gd_3Fe_5O_{12}$ sputtering targets were purchased from Kejing, LTD. The crystalline structure of films is studied by using X-ray diffraction (XRD). The clearly resolved GSGG (444) and GdIG (444) peaks together with Laue oscillation demonstrate the high crystal quality and fully strained GdIG films.

Atomic-scale STEM images were acquired by using a FEI Titan Cubed Themis G2 300 electron microscope with capability of achieving a spatial resolution better than 0.6 Å. This instrument is also equipped with an Energy Dispersive X-ray Spectroscopy (EDXS), which enables element mapping. Z contrast imaging was conducted in high-angle annual dark field (HAADF) mode. The cross-sectional TEM specimen along the desired zone axes was prepared by Zeiss Auriga focused ion beam system. In order to protect GdIG film from being damaged by $Ga^+$ ion beam during ion milling, we have deposited 10 nm Au layer and 1μm polycrystalline Pt layer on the surface of GdIG film. Finally, the cross-sectional TEM specimen was milled by $Ar^+$ ion beam (PIPSⅡ, Gatan) to eliminate the damaged surface layer induced from previous $Ga^+$ ion milling, with an acceleration voltage of 2 kV, 1mA for a few minutes.

Magnetometry measurements at different temperatures were done by using a Quantum Design Magnetic Properties Measurement System (MPMS). Devices were patterned by using standard photolithograph and ion milling. SMR and SOT switching experiments were done by using standard lock-in technique (SR830) in combination with Keithley 6221 current source, together with an electromagnet (EM-5 from East Changing) and electrical transport cryostat (4.2 K – 325 K, SRDK-305D, Sumitomo). Anisotropic field ($H_k$) is also estimated from the planar Hall effect measurements with magnetic fields applied in the sample plane along the current direction. The strength of the interfacial DMI is estimated using a SOT magnetometry method via accessing a Cryogenic triple-axis superconducting magnet (5T-2T-2T). Damping coefficient $\alpha$ is acquired by performing spin-torque ferromagnetic resonance experiment, in which a Keysight MXG analog signal generator (N5183B) is utilized.

The X-ray absorption spectroscopy (XAS) and X-ray magnetic circular dichroism (XMCD) measurements at Fe $L_{2,3}$ and Gd $M_{4,5}$ edges were carried out at the BOREAS beamline of the ALBA synchrotron, Spain using the fully circularly polarized X-ray beam produced by an apple-II type undulator. Magnetic fields were generated collinearly with the incoming X-ray direction by a superconducting vector cryomagnet (Scientific Magnetics) and normal to the sample plane, thus enabling the perpendicular magnetization to be measured. To obtain the spin averaged XAS and the XMCD signals, the absorption signals were measured as a function of the photon energy with directions of photon spin parallel $\mu_+(E)$ and antiparallel $\mu_-(E)$ with the perpendicular magnetization of GdIG samples.

Polar MOKE imaging experiments were done by using a commercial MOKE microscope from evico magnetics with a spatial resolution approaching 500 nm. This microscope was also modified for *in situ* feeding nanosecond voltage pulses (Quantum composer model 5445) both at room temperature and at low temperatures (optical imaging cryostat). The temperature difference between optical imaging cryostat and electrical transport cryostat is calibrated to be within $\pm 0.5$ K for ensuring accurate determination of magnetization and angular momentum compensation temperatures. Nucleation of DWs were made: (1) saturating samples using positive/negative fields, (2) gradually decreasing field to the opposite direction until DWs appears in the racetrack. After DWs become stationary in the track, nanosecond voltage pulses were applied. Velocity is subsequently determined by measuring displacements of DWs in polar MOKE images acquired before and after applying pulses. The current density is computed based on the applied voltage, resistivity measured at different temperatures.

Micromagnetic simulations have been performed by considering two coupled LLG equations including the damping-like torque from the spin-Hall contribution. Further details are given in the Supplementary Materials.

**Acknowledgements**. Work carried out at Tsinghua University was supported by the Basic Science Center Project of NSFC (Grant No. 51788104), National Key R&D Program of China (Grant Nos. 2017YFA0206200 and 2016YFA0302300), the National Natural Science Foundation of China (Grant No. 11774194, 51831005, 1181101082, 11804182), Beijing Natural Science Foundation (Grant No. Z190009), Tsinghua University Initiative Scientific Research Program and the Beijing Advanced Innovation Center for Future Chip (ICFC). M.V. was supported by MINECO FIS2016-78591-C3-2-R (AEI/FEDER, UE). O. T. acknowledges support by the RIEC Nationwide Collaborative Research Project. S. K. was supported by Young Investigator Grant (YIG) from Korean-American Scientists and Engineers Association (KSEA) and Research Council Grant URC-19-090 of the University of Missouri. G.F. and M. C. would like to acknowledge the contribution of the COST Action CA17123 "Ultrafast opto-magneto-electronics for non-dissipative information technology". Authors wish to thank Vélez Saül, Guoqiang Yu, Wei Han, Jiang Xiao and Gerrit Bauer for fruitful discussions.

**Author contributions**

W.J. conceived the idea and designed the experiments. H.Z., Y.D. and T.X. fabricated the thin film, patterned the devices and did the spin transport measurements. H.Z., Y.D., T.X., L.Z. and W.J. performed polar MOKE imaging experiments and data analysis. Y.B., H.Z. and Y.Z. did the AFM experiments. K.X. and X.Z. did the STEM experiments. O.T. and S.K. did the theoretical calculation. H.Z., Y.D., T.X., P.G. and M.V. did the XAS and XMCD measurements. G.F and M.C. developed the micromagnetic model. L.S.T. performed micromagnetic simulations. W.J. wrote the manuscript with inputs from all authors.

**Additional information.**

Supplementary information is available in the online version of the paper. Preprints and permission information is available online at www.nature.com/reprints. Correspondence and requests for materials should be addressed to W.J.

**Competing financial interests**

Authors declare no competing financial interests.

# Figure Captions

**FIG. 1. Magnetization compensation in epitaxial Gd$_3$Fe$_5$O$_{12}$ films.** (A) is the high-resolution STEM image showing atomically sharp interface of GSGG/GdIG bilayer. (B) is the hysteresis loop (M-H) measured perpendicular to the sample plane at different temperatures, showing the presence of PMA and $M_s$ = 33 emu/cc (300 K). (C) is the SMR measured in the Hall geometry at different temperatures in a standard Hallbar device made of GSGG/GdIG/Pt trilayer. The sign reversal of $R_{xy}^{SMR-AHE}$ determines the magnetization compensation temperature $T_M \approx$ 220 K. The vanishing saturation magnetization and increasing coercive fields $H_c$ in the vicinity of $T_M \sim$ 220 K are also evident in (D) and (E). (F) – (I) are the corresponding element-specific XMCD magnetic hysteresis loops for Fe and Gd as a function of $H_z$, measured above and below $T_M$, respectively.

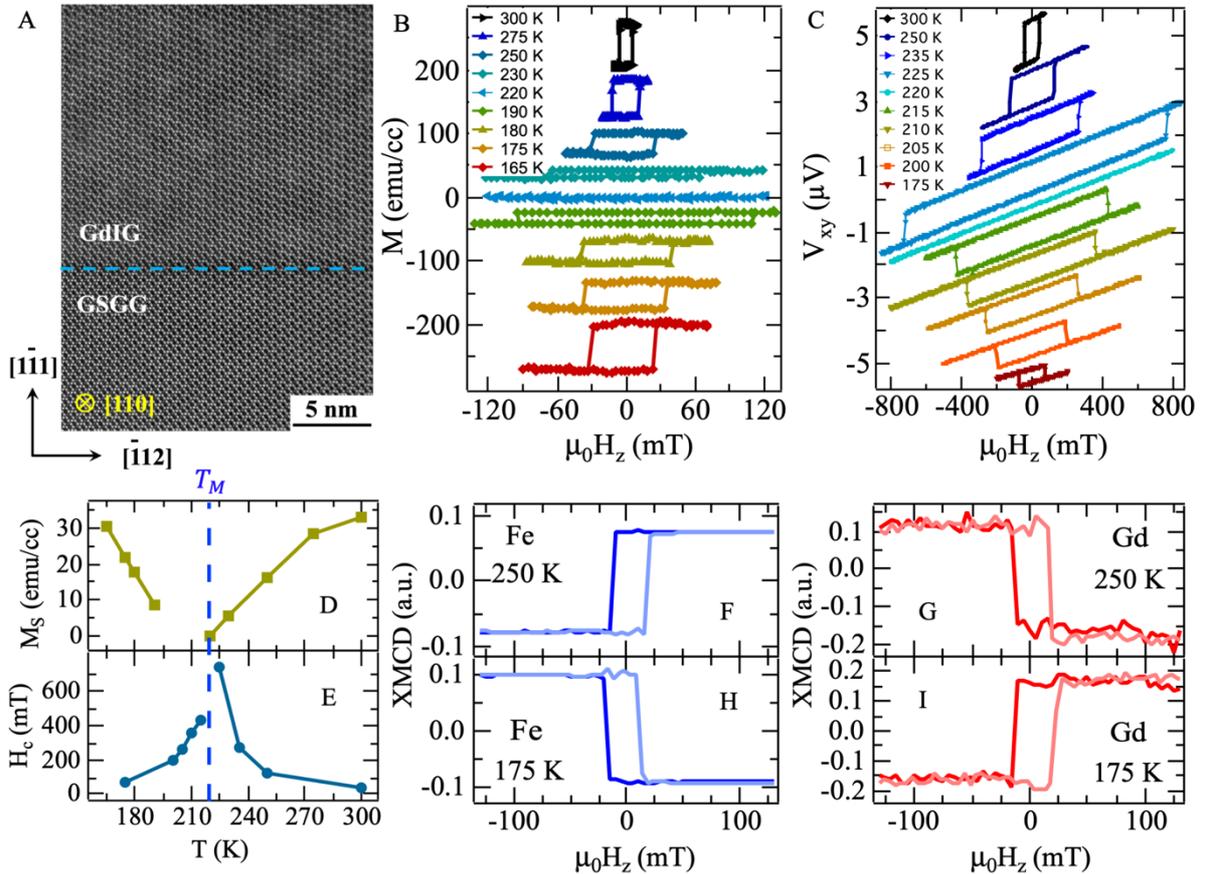

**FIG. 2. Ultrafast DW motion for magnetic fields applied perpendicular to the sample plane ($H_z$) at room temperature.** (A) is an optical image of racetrack-like device made of (15nm)Gd$_3$Fe$_5$O$_{12}$/(4nm)Pt bilayer. (B) – (C) are series of MOKE images acquired after applying sequential pulse currents $j_e = \pm 1.08 \times 10^8$ A/cm$^2$ (28 ns) at $H_z = \pm 8$ mT, yielding an average velocity $V = 50$ m/s (scale bar is 10 µm). Based on the symmetry of $\vec{H}_{sh}$ acting on the Néel type of DWs, the contraction of domain suggests the presence of achiral DWs, as schematically illustrated in the top panel of (C). The orange arrow denotes the magnetization orientation of Fe, the blue arrow is for Gd. (D) are the fast motion of DWs after applying pulse current $j_e = 1.08 \times 10^8$ A/cm$^2$ of duration 34 ns, in which velocity of $V = 2630 \pm 30$ m/s is estimated for $H_z = +8$ mT and $V = 2950 \pm 30$ m/s for $H_z = -8$ mT, respectively. (E) corresponds to DW velocity as a function of pulse durations and amplitudes, at $H_z = +8$ mT. (F) shows DW velocity as a function of pulse duration at $H_z = \pm 9$ mT and $j_e = +1.08 \times 10^8$ A/cm$^2$. (G) is the evolution of DW velocity as a function of $H_z$ at $j_e = +1.08 \times 10^8$ A/cm$^2$, a nonlinear increase of DW velocity following an increase of $|H_z|$ and pulse durations are identified.

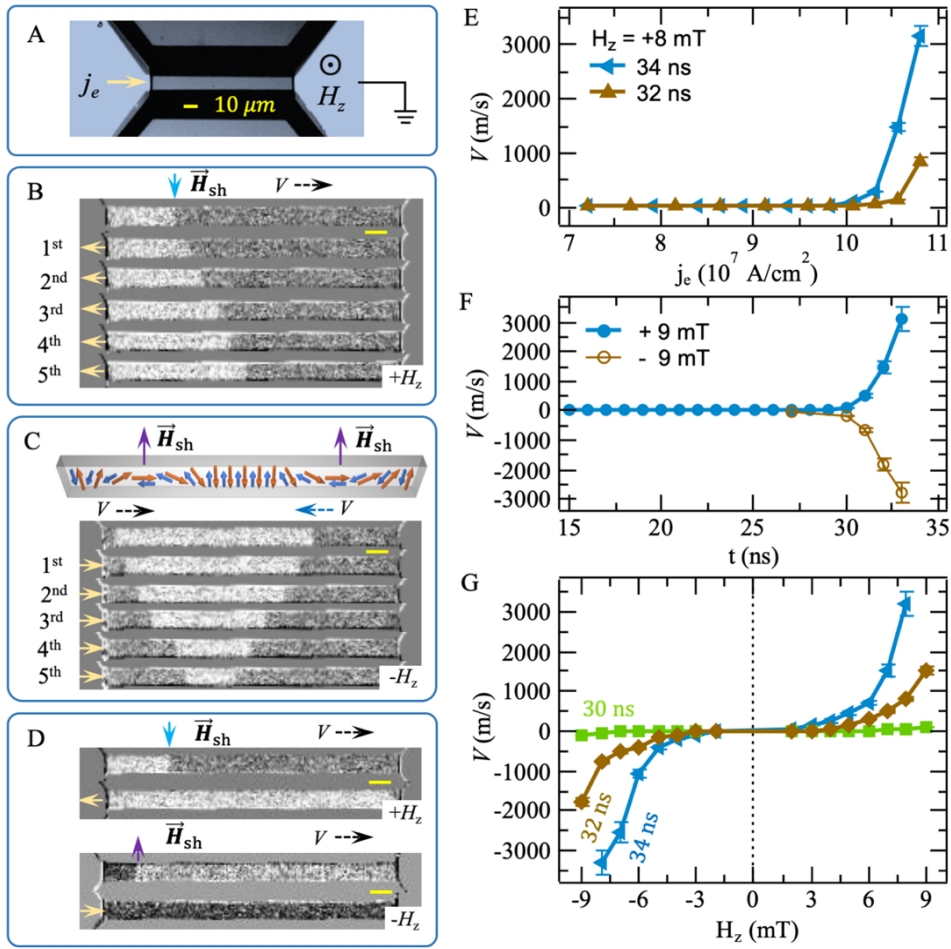

**FIG. 3. DW motion for magnetic fields applied in the sample plane ($H_{x,y}$) at room temperature.** Summarized in (A) are DW velocities as a function of opposite in-plane fields ($\pm H_x$) and opposite polarity of pulse currents ($\pm j_e$). Following the reversed sign of $\pm H_x$ and $\pm j_e$, the direction of DW motion is reversed accordingly (scale bar is 10 μm). Corresponding MOKE imaging data acquired at $\pm H_x = 40$ mT, before and after applying pulse electron current of amplitude $\pm j_e = 1.08 \times 10^8$ A/cm² (duration 25 ns) were given in the top and bottom panels accordingly, which yield a DW velocity $V \approx 3550 \pm 50$ m/s. The direction of DW motion is schematically illustrated in the inset based on the symmetry of $\vec{H}_{sh}$. Presented in (B) is a monotonically increase of DW velocity as a function of pulse duration with $j_e = 1.08 \times 10^8$ A/cm² and $H_x = $ -10 mT. (D) is the evolution of DW velocity a function of pulse amplitudes with a fixed duration of 25 ns in the presence of $H_x = 50$ mT. (D) is the response of Bloch DWs to pulse current ($j_e = 1.08 \times 10^8$ A/cm², 18 ns) in the presence of $H_y$, in which no motion is identified.

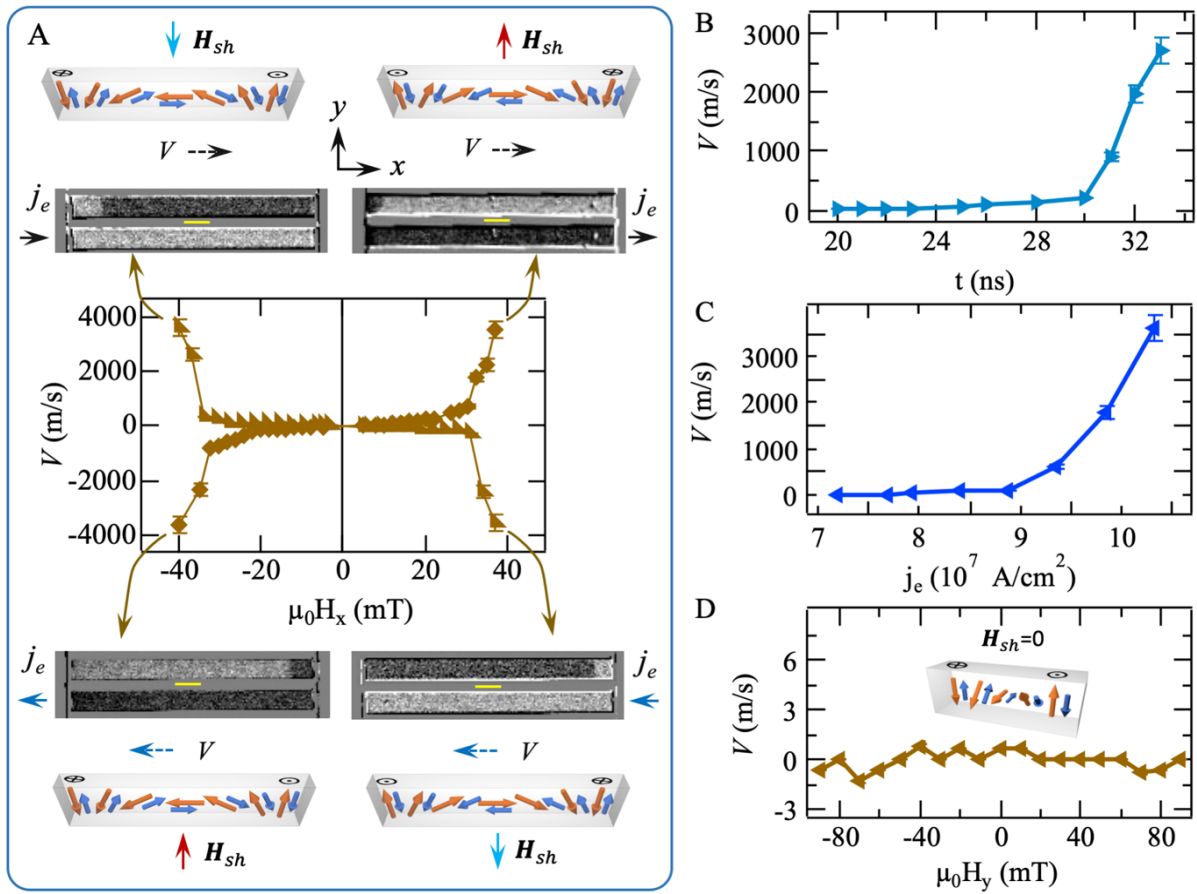

**FIG. 4. Evolution of DW velocity as a function of temperatures.** For a fixed pulse current density $j_e = 1 \times 10^8$ A/cm$^2$ (33 ns), the maximum DW velocity ($V \approx 3500 \pm 300$ m/s) occurs at $T = 230$ K that corresponds to spin angular momentum compensation temperature $T_A$. Due to the large variation of depinning fields in the measured temperature range, perpendicular magnetic fields ($H_z$) were normalized by the corresponding coercive field ($H_c$) at each temperature. Note that this set of data were acquired in an optical cryostat. Shown in (B) is a prototypical DW racetrack devices made of (15nm)Gd$_3$Fe$_5$O$_{12}$/(4nm)Pt bilayer. The position of DWs can be manipulated by passing pulse currents and readout by recording the accompanied SMR voltages, in which the high/low SMR voltages ($V_{xy}$) can be used to presenting 1/0, respectively.

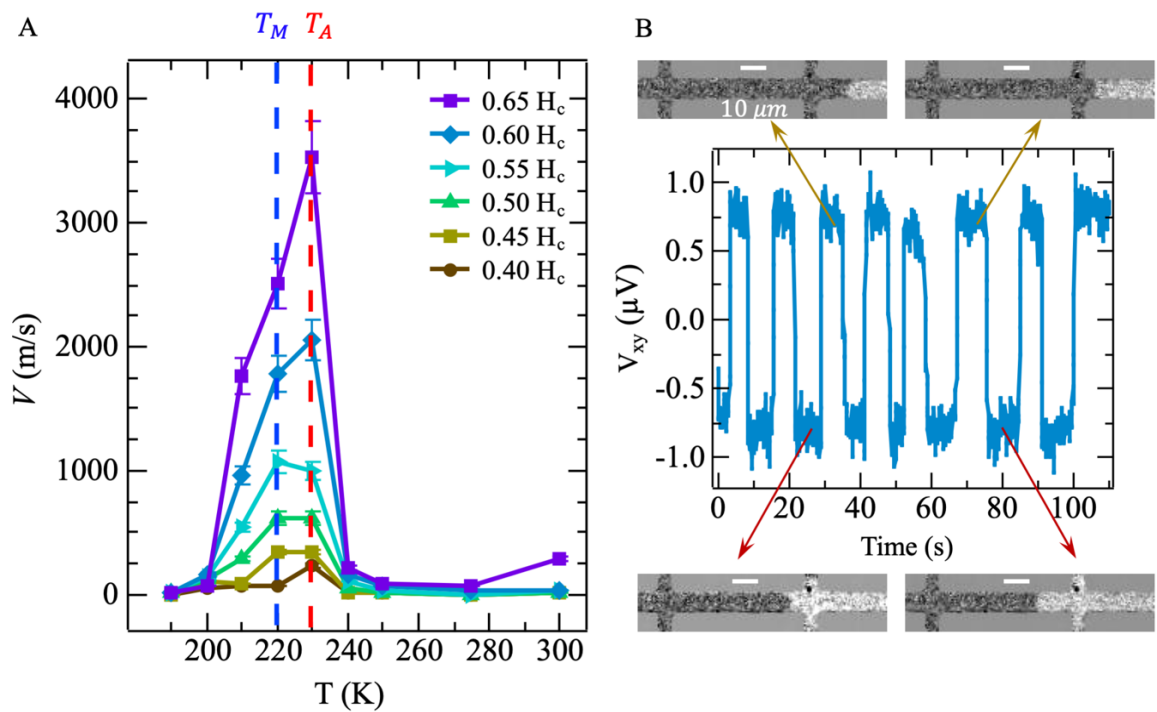